

\newcount\dflag
\dflag = 0


\def\monthname{\ifcase\month
\or Jan\ \or Feb\ \or Mar\ \or Apr\ \or May\ \or June\ %
\or July\ \or Aug\ \or Sept\ \or Oct\ \or Nov\ \or Dec\ 
\else{???}\fi}




\def\endignore{}
\def\ignore #1\endignore{}


\global\hsize = 6.1in
\global\hoffset = .2in
\global\baselineskip = 1.2\baselineskip
\global\parskip = 4pt plus 0.3pt
\global\nulldelimiterspace = 0pt

\predisplaypenalty 5000


\overfullrule=0pt%

\font\titlerm = cmr10 scaled\magstep 4
\font\titlerms = cmr7 scaled\magstep 4
\font\titlermss = cmr5 scaled\magstep 4
\font\titlei = cmmi10 scaled\magstep 4
\font\titleis = cmmi7 scaled\magstep 4
\font\titleiss = cmmi5 scaled\magstep 4
\font\titlesy = cmsy10 scaled\magstep 4
\font\titlesys = cmsy7 scaled\magstep 4
\font\titlesyss = cmsy5 scaled\magstep 4
\font\titleit = cmti10 scaled\magstep 4

\def\titlefont{\def\rm{\fam0\titlerm}
\def\it{\fam\itfam\titleit}
\textfont0 = \titlerm
\scriptfont0 = \titlerms
\scriptscriptfont0 = \titlermss
\textfont1 = \titlei
\scriptfont1 = \titleis
\scriptscriptfont1 = \titleiss
\textfont2 = \titlesy
\scriptfont2 = \titlesys
\scriptscriptfont2 = \titlesyss
\textfont\itfam = \titleit
\rm}

\def\sectionfont{\def\rm{\fam0\tenrm}
\def\it{\fam\itfam\tenit}
\def\bf{\fam\bffam\tenbf}
\textfont0 = \tenrm
\scriptfont0 = \sevenrm
\scriptscriptfont0 = \fiverm
\textfont1 = \teni
\scriptfont1 = \seveni  \scriptscriptfont1=\fivei
\textfont2 = \tensy
\scriptfont2 = \sevensy
\scriptscriptfont2 = \fivesy
\textfont\itfam = \tenit
\textfont\bffam = \tenbf
\rm}

\font\littlefont = cmr7


\def\endid{}
\def\id#1\endid{\number\day\ \monthname \number\year
\hfill #1}

\def\endtitle{}
\def\title#1\endtitle{\vskip.15in\titlefont
\global\baselineskip = 2\baselineskip
#1\vskip.3in
\baselineskip = 0.5\baselineskip\sectionfont}

\def\endauthors{}
\def\authors#1\endauthors{
#1\if\dflag = 0
\fi}

\def\endabstract{}
\def\abstract#1\endabstract{\vskip .2in%
\centerline{\sectionfont\bf Abstract}%
\vskip .1in%
\noindent#1%
\footline = {\hfil}\pageno = 0
\vfill\eject
\pageno = 1\footline{\centerline{\sectionfont\folio}}
}


\newcount\nsection
\newcount\nsubsection
\newcount\nappendix

\nsection = 0
\nsubsection = 0
\nappendix = 0

\newcount\nequation
\nequation = 1
\def\eqlabel{(1.1)}

\def\secname{\ifcase\nappendix%
\number\nsection
\or{A}\or{B}\or{C}\or{D}\or{E}\or{F}\or{G}\or{H}\or{I}\or{J}\or{K}\or{L}%
\or{M}\or{N}\or{O}\or{P}\or{Q}\or{R}\or{S}\or{T}\or{U}\or{V}\or{W}\or{X}%
\or{Y}\or{Z}
\else{?}
\fi}

\def\section#1{%
\advance\nsection by 1
\nequation = 1
\xdef\eqlabel{({\rm\secname}.1)}
\nsubsection = 0
\nequation = 1
\bigbreak\noindent
\centerline{\sectionfont\bf\number\nsection.\ #1}
\nobreak\smallskip\sectionfont\nobreak\par\nobreak}

\def\subsection#1{\advance\nsubsection by 1
\bigbreak\noindent
\centerline{\sectionfont\it\secname.\number\nsubsection.\ #1}%
\nobreak\smallskip\rm\nobreak\par\nobreak}

\def\appendix#1{%
\advance\nappendix by 1
\xdef\eqlabel{({\rm\secname}.1)}
\nsubsection = 0
\nequation = 1
\bigbreak\noindent
\centerline{\sectionfont\bf Appendix\ \secname.~#1}
\nobreak\smallskip\rm\nobreak\par\nobreak}


\newcount\nref
\global\nref = 1

\def\ref#1#2{\xdef #1{[\number\nref]}
#1
\ifnum\nref = 1\global\xdef\therefs{\noindent[\number\nref] #2\ }
\else
\global\xdef\oldrefs{\therefs}
\global\xdef\therefs{\oldrefs\vskip.1in\noindent[\number\nref] #2\ }%
\fi%
\global\advance\nref by 1
}

\def\aref#1{[\number\nref]
\ifnum\nref = 1\global\xdef\therefs{\noindent[\number\nref] #1\ }
\else
\global\xdef\oldrefs{\therefs}
\global\xdef\therefs{\oldrefs\vskip.1in\noindent[\number\nref] #1\ }%
\fi%
\global\advance\nref by 1
}

\def\listrefs{\vfill\eject\section{References}\therefs}


\newcount\cflag

\def\eqname{({\rm\secname}.\number\nequation)}

\def\nexteqno{\ifnum\cflag = 0
\global\advance\nequation by 1
\fi
\global\cflag = 0
\xdef\eqlabel{\eqname}}

\def\lasteqno{\global\advance\nequation by -1
\xdef\eqlabel{\eqname}}

\def\label#1{\xdef #1{\eqname}
\ifnum\dflag = 1
{\escapechar = -1
\xdef\draftname{\littlefont\string#1}}
\fi}

\def\clabel#1#2{\xdef\eqlabel{({\rm\secname}.\number\nequation #2)}
\global\cflag = 1
\xdef #1{\eqlabel}
\ifnum\dflag = 1
{\escapechar = -1
\xdef\draftname{\string#1}}
\fi}

\def\cclabel#1#2{\xdef\eqlabel{(#2)}
\global\cflag = 1
\xdef #1{\eqlabel}
\ifnum\dflag = 1
{\escapechar = -1
\xdef\draftname{\string#1}}
\fi}


\def\eeq{}

\def\eqnn #1\eeq{$$ #1 $$}

\def\eq #1\eeq{\xdef\draftname{\ }
$$ #1
\eqno{\eqlabel \rlap{\ \draftname}} $$
\nexteqno}



\def\eol{& \eqlabel \rlap{\ \draftname} \crcr
\nexteqno
\xdef\draftname{\ }}

\def\eeol{& \eqlabel \rlap{\ \draftname}
\nexteqno
\xdef\draftname{\ }}

\def\eolnn{\cr
\global\cflag = 0
\xdef\draftname{\ }}


\def\eqa #1\eeq{\xdef\draftname{\ }
$$ \eqalignno{ #1 } $$
\global\cflag = 0}


\newcount\nfig
\global\nfig = 1

\def\fg#1\efig{\vskip .5in\noindent Fig.\ \number\nfig:\ #1%
\global\advance\nfig by 1}


\def\ie{{\it i.e.\/}}
\def\eg{{\it e.g.\/}}

\def\etc{{\it etc}}
\def\etal{{\it et.\ al.\/}}



\newcount\jstyle
\jstyle = 0

\def\jref#1#2#3#4{%
\ifnum \jstyle = 0
{\it #1} {\bf #2}, #3 (#4)
\else
{\it #1} #2\ (#4)\ #3
\fi}

\def\AP#1#2#3{\jref{Ann.\ Phys.}{#1}{#2}{#3}}

\def\NPB#1#2#3{\jref{Nucl.\ Phys.}{B#1}{#2}{#3}}

\def\PLB#1#2#3{\jref{Phys.\ Lett.}{#1B}{#2}{#3}}
\def\PR#1#2#3{\jref{Phys.\ Rep.}{#1}{#2}{#3}}
\def\PRD#1#2#3{\jref{Phys.\ Rev.}{D#1}{#2}{#3}}

\def\PRL#1#2#3{\jref{Phys.\ Rev.\ Lett.}{#1}{#2}{#3}}

\def\PRV#1#2#3{\jref{Phys.\ Rev.}{#1}{#2}{#3}}


\def\to{\mathop{\rightarrow}}


\def\myint{\int\mkern-5mu}
\def\frac#1#2{{{#1} \over {#2}}\,}  


\def\Dsl{\hbox{\kern.1em/\kern-.7000em$D$}} 



\def\scr#1{{\cal #1}}

\def\mybar#1{\kern 0.8pt\overline{\kern -0.8pt#1\kern -0.8pt}\kern 0.8pt}
\def\sla#1{\raise.15ex\hbox{$/$}\kern-.57em #1}
\def\Sla#1{\kern.15em\raise.15ex\hbox{$/$}\kern-.72em #1}

\def\roughly#1{\mathrel{\raise.3ex\hbox{$#1$\kern-.75em%
    \lower1ex\hbox{$\sim$}}}}


\def\tr{\mathop{\rm tr}}

\def\Im{\mathop{\rm Im}}


\def\bra#1{\langle #1 |}
\def\ket#1{| #1 \rangle}
\def\braket#1#2{\langle #1 | #2 \rangle}

\def\avg#1{\langle #1 \rangle}

\def\Avg#1{\left\langle #1 \right\rangle}



\hyphenation{ano-ma-ly ano-ma-lies}
\hyphenation{ba-ry-on ba-ry-ons}
\hyphenation{la-gran-gian la-gran-gians}
\hyphenation{phy-sics phy-si-cal}

\def\al{\alpha}
\def\del{\delta}
\def\Del{\Delta}
\def\gam{\gamma}

\def\ep{\epsilon}

\def\Lam{\Lambda}
\def\om{\omega}

\def\sig{\sigma}

\def\ChPT{\raise.45ex\hbox{$\chi$}PT}

\def\rhs{right-hand side}

\def\hc{{\rm h.c.}}


\def\MeV{{\rm \ MeV}}
\def\GeV{{\rm \ GeV}}

\def\LQCD{\Lam_{\rm QCD}}

\def\emd{electromagnetic mass difference}
\def\pid{$\pi^+$--$\pi^0$ mass difference}


\vskip -0.2in
\id
HUTP-95/A006, MIT-CTP-2412
\endid

\title
\centerline{Heavy Meson Electromagnetic}
\centerline{Mass Differences from QCD}
\endtitle

\authors
\centerline{Markus A. Luty\footnote{$^*$}
{e-mail: luty@ctp.mit.edu}}
\vskip .1in
\centerline{\it Center for Theoretical Physics}
\centerline{\it Massachusetts Institute of Technology}
\centerline{\it Cambridge, Massachusetts 02139}
\vskip .1in
\centerline{Raman Sundrum\footnote{$^\dagger$}
{e-mail: sundrum@huhepl.harvard.edu}}
\vskip .1in
\centerline{\it Lyman Laboratory of Physics}
\centerline{\it Harvard University}
\centerline{\it Cambridge, Massachusetts 02138}
\endauthors

\abstract
We compute the electromagnetic mass differences of mesons containing a
single heavy quark in terms of measurable data using QCD-based arguments in
heavy-quark effective theory.
We derive an unsubtracted dispersion relation that shows that the mass
differences are calculable in terms of the properties of the lowest-lying
physical intermediate states.
We then consider the problem in the large-$N$ limit, where $N$ is the number
of QCD colors.
In this limit, we can write a kind of double-dispersion relation for the
amplitude required to determine the electromagnetic mass difference.
We use this to derive analogs of the Weinberg sum rules for heavy meson
matrix elements valid to leading order in $1/N$ and to $O(1/m_Q)$ in the
heavy quark expansion.
In order to obtain our final result, we assume that the electromagnetic mass
differences and sum rules are dominated by the lowest-lying states in analogy
with the situation for the $\pi^+$--$\pi^0$ mass difference.
Despite the fact that some of the matrix elements appearing in our final result
have not yet been accurately measured, we can obtain useful estimates:
for example, we obtain $(M_{B^+} - M_{B^0})^{\rm EM} \simeq +1.8 \MeV$.
We argue that our results are accurate to about $30\%$.
\endabstract


\section{Introduction}
The computation of the \emd s of hadrons is one of the classic problems in
strong interaction physics (see \eg\
\ref\oldbarry{See \eg\ R. P. Feynman and G. Speisman, \PRV{94}{500}{1954};
M. Cini, E. Ferrari, and R. Gatto, \PRL{2}{7}{1959}.}
\ref\oldpi{T. Das, G. S. Guralnik, V. S. Mathur, F. E. Low, and J. E. Young,
\PRL{18}{759}{1967}; T. Das, V. S. Mathur, S. Okubo, \PRL{19}{859}{1967}. }
\ref\cott{W. N. Cottingham, \AP{25}{424}{1963}};
for reviews, see
\ref\Feynman{R. P. Feynman, {\it Photon--Hadron Interactions} (Benjamin,
1972).}
\ref\Zee{A. Zee, \PR{3}{129}{1972}.}
\ref\newpi{S. Pokorski, {\it Gauge Field Theories} (Cambridge, 1987).}).
At one time, it was believed that electromagnetism was the only source of
isospin breaking, so that computations of \emd s could be compared directly
with experiment.
With the advent of QCD, it is now understood that isospin breaking arises
both from electromagnetism and from differences in the $u$ and $d$ quark
current masses.
The modern motivation for computing \emd s is to disentangle the
electromagnetic contributions to isospin-violating mass differences from
those of the current quark masses in order to obtain information about the
current quark masses.\footnote{$^*$}
{The \pid\ is a special case, since quark masses do not contribute to it at
leading order.
Therefore, the electromagnetic contribution is expected to dominate, and it
can be compared directly with experiment.}

In this paper, we will compute the $O(e^2)$ electromagnetic mass differences of
lowest-lying mesons with quantum numbers $Q\mybar u$ and $Q\mybar d$ (denoted
here by $P_{\mybar\ell}$ ($\ell = u$ or $d$).
We work to $O(1/m_Q)$ in the heavy-quark expansion and to leading order in the
$1/N$ expansion, where $N$ is the number of QCD colors.
We give a detailed exposition of the formalism used and give useful estimates
of the \emd s.\footnote{$^\dagger$}
{The \emd s of heavy mesons have also been considered in
ref.~\ref\Goity{J. L. Goity, \PLB{303}{337}{1993}.},
but the methods used are not based on a systematic approximation of QCD.}
A detailed comparison to experiment and the extraction of information about
the light quark masses is carried out in a separate paper
\ref\us{M. A. Luty and R. Sundrum, in preparation.}.

We follow as closely as possible the method of the classic calculation of the
\pid\ \oldpi\newpi.
The basic strategy is to write the $O(e^2)$ self-energy of the meson in
terms of the forward Compton amplitude $T$, and then write a dispersion
relation for $T$ expressing it in terms of measurable data.
In the case of the \pid, one can use the fact that the pions are
pseudo-Nambu--Goldstone bosons to write $T$ in terms of {\it vacuum}
current correlation functions depending only on a single kinematic invariant.
Dispersion relations then relate these to data measured in
$e^+ e^- \to {\rm hadrons}$ and $\tau$ decays.
For heavy mesons we must work directly with $T$, which depends on two
kinematic invariants.
This makes the computation of the heavy meson \emd s more complicated than
the \pid.

We begin in section 2 by writing an unsubtracted dispersion relation that
shows that $T$ is determined by the properties of low-lying meson states and
the lowest-lying excitations of the heavy--light quark system.
Unfortunately, this dispersion relation cannot be used directly to compute
$T$;
one reason is that it depends on structure functions for timelike photon
momenta, which are not measurable in practice.
In section 3, we consider the problem in the combined heavy-quark and
large-$N$ limits.
In the large-$N$ limit, we can write $T$ in terms of heavy-meson form factors
and meson scattering amplitudes.
We derive sum rules relating the properties of the masses and matrix elements
of the states that determine $T$.
(These are exact analogs of the Weinberg sum rules for vector- and
axial-vector correlation functions in the vacuum
\ref\Weinsum{S. Weinberg, \PRL{18}{507}{1967};
C. Bernard, A. Duncan, J. Lo Secco, and S. Weinberg,
\PRD{12}{792}{1975}.}.)
By assuming that these sum rules are approximately saturated by the
lowest-lying states, we can compute the heavy meson \emd s in terms of
heavy meson form factors.
This is analogous to the successful classic calculation of the \pid.
In section 4, we consider the $1/m_Q$ corrections to these results, and section
5 summarizes our results and gives our conclusions.
This is a rather long paper, but the reader can get the main ideas by reading
sections 2 and section 3 through subsection 3.1, followed by section 5 (which
contains a summary of the main results).
The sections omitted in this way consist mainly of repeated application of the
ideas in the first part of the paper.

Our final result is similar to the prescription used long ago to compute
baryon \emd s \oldbarry, but we emphasize that this prescription was never put
on a firm foundation \Feynman\Zee.
In fact, it is ironic that we are not able to extend our results to baryons
because the large-$N$ limit is more complicated for baryons.

\section{Technical Preliminaries}
In this section, we derive some technical results that provide the foundation
for the rest of the paper.
Although we focus on the electromagnetic mass differences of heavy mesons in
this paper, most of the formalism in this section can be applied to any
type of hadron.

\subsection{Renormalization and Finiteness}
The $O(e^2)$ electromagnetic contribution to the mass difference of the
pseudoscalar mesons $P_{\mybar\ell}$ ($\ell = u, d$) with flavor quantum
numbers $Q\mybar\ell$ is given by
\eq
\label\emass
\eqalign{
\Del M \equiv M_{\mybar u} - M_{\mybar d} =
\frac{ie^2}{2} &\myint \frac{d^4 q}{(2\pi)^4}\,
\frac {\Del T(p, q)}{q^2 + i0+} \cr
&- \bra{P_{\mybar u}(p)} \del\scr L \ket{P_{\mybar u}(p)}
+ \bra{P_{\mybar d}(p)} \del\scr L \ket{P_{\mybar d}(p)}, \cr}
\eeq
where $\del\scr L$ is the $O(e^2)$ counterterm in the underlying lagrangian
required to render the result finite (see below);
$M$ is the mass of the heavy mesons in the absence of electromagnetic
interactions;
we have defined
\eq
\label\DelTdef
\Del T \equiv T_{\mybar u} - T_{\mybar d},
\eeq
where
\eq
\label\Tdef
T_{\mybar\ell}(p, q) \equiv i\myint d^4 x\, e^{iq\cdot x}
\bra{P_{\mybar\ell}(p)} T J^\mu(0) J_\mu(x) \ket{P_{\mybar\ell}(p)}.
\eeq
Here, $J^\mu$ is the electromagnetic current, so $T$ is a trace over the
forward Compton amplitude for scattering of photons from heavy mesons.
For states containing heavy mesons, we use the normalization
\eq
\braket{P(p)}{P(p')} = (2\pi)^3 \del^3(\vec p - \vec p\,')
\eeq
appropriate for the heavy particle effective theory.
Similar equations hold for the lowest-lying heavy-light vector mesons $P^*$
that are related to $P$ by heavy-quark symmetry.

We now consider the counterterm $\del\scr L$ in eq.~\emass.
Because we will use the heavy-quark expansion, the underlying lagrangian is
that of heavy quark effective theory
\aref{H. Georgi, \PLB{240}{447}{1990}.}.
To $O(1/m_Q)$ in the heavy-quark expansion, the lagrangian is
\eq
\label\theL
\eqalign{
\scr L = &-\frac 14 F^{\mu\nu} F_{\mu\nu}
- \frac 14 \tr(G^{\mu\nu} G_{\mu\nu}) + \mybar q i\Dsl q
+ \mybar Q iv\cdot D Q \cr
&\quad - \frac 1{2m_Q} \mybar Q D^2 Q
+ \frac{g_{\rm s} a}{4m_Q} \mybar Q \sig^{\mu\nu} G_{\mu\nu} Q
+ \frac{e_Q b}{4m_Q} F^{\mu\nu} \mybar Q \sig_{\mu\nu} Q
+ \cdots. \cr}
\eeq
Here, $F^{\mu\nu}$ is the electromagnetic field strength, $G^{\mu\nu}$ is
the gluon field strength and $v$ is the $4$-velocity of the heavy meson.
We set the light current quark masses to zero for our computation, since we
are not interested in $O(e^2 m_{u,d,s})$ effects.
The coefficient of the term $\mybar Q D^2 Q$ is fixed by reparameterization
invariance
\ref\reparam{M. Luke and A. V. Manohar, \PLB{286}{348}{1992}.};
the coefficients $a$ and $b$ can be
fixed by matching in QCD perturbation theory and are unity to leading order
in $\alpha_s(m_Q)$
\ref\match{A. Falk, B. Grinstein, and M. Luke, \NPB{357}{185}{1991}.}.

All divergences that appear when computing physical quantities to $O(1/m_Q)$
can be absorbed into counterterms of the form appearing in eq.~\theL.
This severely restricts the ultraviolet behavior of the matrix elements that
appear in eq.~\emass, since only isospin-violating counterterms can
contribute to the mass difference.
Since we are neglecting light current quark masses, the only isospin-violating
term in the lagrangian is the light-quark kinetic term $\bar q i\Dsl q$.
(It violates isospin because the light quarks have different charges.)
However, matrix elements of $\bar q i\Dsl q$ between physical states vanish
by the equations of motion.
Thus, for the $Q\mybar u$--$Q\mybar d$ mass difference, there is no
counterterm contribution, and the integral on the \rhs\ of eq.~\emass\ must
converge.
(A similar argument in the context of full QCD is given in
ref.~\ref\Coll{J. C. Collins, \NPB{149}{90}{1979}.}.)
In fact, since the quark charges are arbitrary parameters, the contributions
from $\Del T_{qq}$ and $\Del T_{Qq}$ must individually converge.

The \emd s get contributions from isospin-violating counterterms at order
$1/m_Q^2$, such as
\eq
\del\scr L = \frac c{m_Q^2} F^{\mu\nu} \mybar q \gam_\mu D_\nu q.
\eeq
The integral for $\Del M$ is convergent in full QCD \Coll, so this simply
means that the $O(1/m_Q^2)$ contribution to the \emd\ is sensitive to momenta
up to $\sim m_Q$, but the integrand falls off for momenta above $m_Q$ fast
enough so that the integral converges.

It is convenient to rewrite $\Del T$ in a useful form using some elementary
isospin group theory.
We first split the the electromagnetic current into heavy-quark and
light-quark components
\eq
J^\mu = J_Q^\mu + J_q^\mu.
\eeq
The heavy-quark current is
\eq
J_Q^\mu = \scr Q_Q \mybar Q \gam^\mu Q,
\eeq
where $\scr Q_c = \frac 23$, $\scr Q_b = -\frac 13$.
The light-quark current can be decomposed into isospin 0 and 1 components
\eq
J^\mu_q = J^\mu_{q0} + J^\mu_{q1},
\eeq
where
\eq
J^\mu_{q0} = \frac{\scr Q_u + \scr Q_d}2 \mybar q \gam^\mu q, \qquad
J^\mu_{q1} = \frac{\scr Q_u - \scr Q_d}2 \mybar q \tau_3 \gam^\mu q, \qquad
q \equiv \pmatrix{u \cr d \cr},
\eeq
and $\scr Q_u = \frac 23$, $\scr Q_d = -\frac 13$.
We can then write
\eq
\Del T = \Del T_{qq} + \Del T_{Qq}
\eeq
in terms of the state $\bra P \equiv \bra{P_{Q\mybar u}}$ alone:
\eqa
\label\Tqqdef
\Del T_{qq}(p, q) &\equiv 2i \myint d^4x\, e^{iq\cdot x}
\bra{P(p)} T J^\mu_{q0}(0) J_{q1\mu}(x)
\ket{P(p)} + (q \to -q), \eol
\label\TQqdef
\Del T_{Qq}(p, q) &\equiv 2i\myint d^4x\, e^{iq\cdot x}
\bra{P(p)} T J^\mu_Q(0) J_{q1\mu}(x)
\ket{P(p)} + (q \to -q). \eeol
\eeq
We denote the corresponding contributions to $\Del M$ by $\Del M_{qq}$ and
$\Del M_{Qq}$, respectively.

\def\myvec#1{{\vec{#1}\,}}
\def\qv{\myvec q}

\subsection{Dispersion Relations and Low-energy Dominance}
Eq.~\emass\ reduces the problem of computing the \emd s to the problem of
determining the forward Compton amplitude $\Del T$ defined in eq.~\DelTdef.
We address this problem by writing a dispersion relation that expresses
$\Del T$ in terms of measurable data.
$\Del T$ is a function of two kinematic invariants, which we take to be
$\qv^2$ and $\nu \equiv q_0$ in the frame where $p = (M_P, \vec 0\,)$.
We can then write a spectral representation for $\Del T$ by inserting a
complete set of states into eq.~\Tdef.
The result is
\eq
\label\disp
\Del T(\nu, \qv^2) =
-\frac 2\pi\int_0^\infty d\nu'\,
\frac{\nu' \Im \Del T(\nu', \qv^2)}{\nu^2 - {\nu'}^2 + i0+},
\eeq
which can be thought of as a fixed-$\qv^2$ dispersion relation.
Dispersion relations such as this in general require subtraction because
of the ultraviolet behavior of products of currents.
However, due to isospin invariance, there is no counterterm that modifies the
product of currents appearing in $\Del T$, so the operation of inserting a
complete set of states that was used to derive eq.~\disp\ is valid without
modification.\footnote{$^*$}
{In more detail:
the operators $J^\mu_{0,1}$ are defined by functional differentiation with
respect to appropriate sources $S^\mu_{0,1}$ added to the lagrangian.
All divergences in this lagrangian can be canceled by counterterms depending
on the sources with dimension 4 or less.
Because there is no counterterm of the form $S_0 S_1$ allowed by isospin
invariance, there is no short-distance singularity in the product of currents
appearing in $\Del T$, and the manipulations used to derive eq.~\disp\ are
valid.
It is interesting to note that it was guessed that eq.~\disp\ did not require
subtraction before the advent of QCD;
see for example ref.~\Feynman.}

$\Im \Del T$ is determined by physical matrix elements.
In the $m_Q \to \infty$ limit, the relation is particularly simple:
\eq
\label\data
\Im \Del T(\nu, \qv^2) = 2\pi \sum_n \del(\nu - \Del_n)
\bra{P(p)} J^\mu_0(0) \ket{n(\qv)}
\bra{n(\qv)} J_{1\mu}(0) \ket{P(p)} + \hc,
\eeq
where $M + \Del_n$ is the mass of the state $n$, and
$J^\mu_0 \equiv J^\mu_{q0} + J^\mu_Q$, $J^\mu_1 \equiv J^\mu_{q1}$.
The sum over $n$ is over a complete set of states with fixed 3-momentum
$\qv$.
This in principle solves the problem of determining $\Del T$ in terms of
physical data, since $\Im \Del T(\nu, q^2)$ is determined by structure
functions for inelastic scattering off the heavy meson.
However, eqs.~\disp\ and \data\ by themselves do not give a practical method of
determining $\Del T$;
for example, experimentally inaccessible data for timelike photon momenta
(``timelike structure functions'') are required on the \rhs\ of eq.~\disp.
We will see in the next section that progress can still be made in the context
of the large-$N$ limit of QCD.

The main use of the dispersion relation eq.~\disp\ for our purposes is that
it makes manifest that the electromagnetic mass differences are insensitive to
the properties of very heavy states.
We have argued that the integral that determines $\Del M$ converges, so
it is dominated by $\Del T(\nu, \qv^2)$ with
$\nu^2 \sim \qv^2 \sim \LQCD^2$.
Now consider the contributions of an intermediate state $n$ to $\Del T$
in this kinematic region:
the delta function in eq.~\data\ means that a state with mass $M + \Del_n$
contributes at $\nu' = \Del_n$ in the \rhs\ of eq.~\disp.
Therefore, the fact that the integral over $\nu'$ in eq.~\disp\ converges
means that the contribution of states with $\Del_n \gg \LQCD$ is suppressed.

The dispersion relation eq.~\disp\ also shows that we can continue the
integral in eq.~\emass\ to Euclidean momenta.
(Specifically, we write $q = (iq_{E0}, \qv)$, so that
$d^4 q = i d^4 q_E$ and $q^2 = -q_E^2$.)
We also note that because the integrand is integrated over $d^4 q_E$, we can
angularly average it in Euclidean momentum space without changing the result.
We then have
\eq
\label\eemass
\Del M = \frac{e^2}{2} \myint \frac{d^4 q_E}{(2\pi)^4}\,
\frac{\avg{\Del T}(q_E^2)}{q_E^2},
\eeq
where $\avg{\Del T}$ is the angular average of $\Del T$.
Eq.~\eemass\ shows that we need not determine the full kinematic
behavior of $\Del T$.
For example, $\avg{\Del T}$ can be shown to be insensitive to the scaling
behavior in the deep-inelastic region, in agreement with the fact that
infinitely many states contribute to scaling behavior \Zee.

This problem of computing hadron \emd s has a long and involved history
\Feynman\Zee, and we will not give a detailed discussion of the literature.
We do wish to remark that one common approach to the problem of baryon \emd s
has been to write a fixed-$q^2$ (as opposed to fixed-$\qv^2$) dispersion
relation and attempt to express the electromagnetic mass differences in terms
of measured {\it spacelike} structure functions \cott\Feynman\Zee.
However, there are good reasons to believe that the fixed-$q^2$ dispersion
relation requires subtraction.
(The necessity of the subtraction is related to the small $x$ behavior of
the structure functions; see \eg\ ref.~\Zee.)
Since the electromagnetic splittings cannot be determined without knowing the
subtraction constant, we do not follow this approach in this paper.
We will, however, make some brief comments about the relation between this
approach and ours in the next section.

\section{$\Del M$ in the Large-$N$ and Heavy-quark Limit}
In this section, we consider the computation of $\Del M$ in the combined
large-$N$ and heavy-quark limit.
We will consider $O(1/m_Q)$ corrections in the next section, but we will not
attempt to go beyond leading order in the $1/N$ expansion in this paper.

For large $N$, QCD reduces to a weakly-coupled field theory with
infinitely many meson fields whose interactions are polynomials in momenta
\ref\Witten{See \eg\ E. Witten, \NPB{160}{57}{1979}.}.
The transition amplitudes can be expanded systematically in powers of $1/N$,
and the leading term in this expansion corresponds to keeping only tree graphs
in the mesonic theory.
The meson graphs that contribute to $\Del T$ are shown in fig.~1.

The representation of $\Del T$ as a sum of tree graphs gives a kind of
double-dispersion relation that will allow us to determine the heavy-meson
\emd s in terms of measurable matrix elements.
For example, it is clear that the graphs with intermediate heavy-meson lines
(the first graph in fig.~1) are related to heavy meson form factors.
However, we must also obtain information about the remaining graphs, which
are not obviously related to form factors.
We will do this by imposing consistency between the hadronic theory and
properties of QCD and heavy-quark effective theory, such as those discussed in
the previous section.
These consistency conditions are expressed in terms of sum rules relating
the (infinitely many) couplings of the states that appear in the graphs.
If we assume that the sum rules are saturated by the lowest-lying states
(consistent with the low-energy dominance proved above), we will find that
the unknown contributions from the graphs in fig.~1 with no heavy-meson
intermediate states give a numerically negligible contribution to the \emd.

For baryons, meson loops are {\it not} suppressed in the large-$N$ limit
(see \Witten
\ref\semiclass{N. Dorey, J. Hughes, and M. Mattis, hep-ph/9406406;
A. V. Manohar, UCSD/PTH 94-14, hep-ph/9407211.}),
so the arguments in this section cannot be simply extended to this case.
Recent progress in the $1/N$ expansion for baryons \semiclass\ may be relevant
to overcoming this difficulty.

We now consider the contributions $\Del M_{qq}$ and $\Del M_{Qq}$
(defined by eqs.~\Tqqdef\ and \TQqdef) in turn.

\subsection{$\Del M_{qq}$}
Consider $\Del M_{qq}$ from the point of view of the large-$N$ mesonic
field theory.
The sum of graphs that gives the amplitude $\Del T_{qq}$ can be written
\eq
\label\dtqq
\Del T_{qq}(\nu^2, q^2) =
-2\sum_n \frac{\Del_n W_n(\nu^2, q^2)}{\nu^2 - \Del_n^2 + i0+} +
2 C(\nu^2, q^2),
\eeq
where the first term is the contribution from the first graph in fig.~1, and
the second term is the sum of the remaining graphs.
Because $\Del T_{qq}$ is even under $q \mapsto -q$, we consider it a function
of $\nu^2$ rather than $\nu$.
The sum on $n$ in the first term runs over the (infinitely many) heavy mesons
with mass $M + \Del_n$ that appear in the intermediate-state propagators in
the first graph of fig.~1.
The pole structure in this term arises by combining the heavy-meson
propagators $1 / (\pm \nu - \Del_n + i0+)$.
The vertices in the large-$N$ limit are polynomials, and so the only
non-polynomial dependence of $W_n$ and $C$ in eq.~\dtqq\ comes from the
meson propagators in figs.~1 and 2, which give poles in $q^2$.
Therefore, $W_n$ and $C$ are polynomial in $\nu^2$.

We can give a physical interpretation of $W_n(\nu^2, q^2)$ by noting that for
$q^2$ spacelike, the only real intermediate states that can appear in
$\Del T_{qq}$ in the large-$N$ limit are single heavy meson states.
{}From eq.~\dtqq\ we compute
\eq
\Im \Del T_{qq}(\nu^2, q^2) = \pi
\sum_n W_n(q^2) \del(\nu - \Del_n),
\eeq
where we have defined $W_n(q^2) \equiv W_n(\Del_n^2, q^2)$.
Substituting this into the spectral representation eq.~\data, we obtain
\eq
\label\wndata
W_n(q^2) = 2 \bra{P(p)} J^\mu_{q0}(0) \ket{n(\qv)}
\bra{n(\qv)} J_{q1\mu}(0) \ket{P(p)} + \hc
\eeq
Therefore, $W_n(q^2)$ is proportional to on-shell heavy-meson form factors.
(We can also see this directly by looking at diagrams.)

Because the functions $W_n$ are polynomial in $\nu^2$, we can expand $W_n$
around $\nu^2 = \Del_n^2$ to obtain
\eq
\label\dtqqtoo
\Del T_{qq}(\nu^2, q^2) =
-2 \sum_n \left[ \frac{\Del_n W_n(q^2)}{\nu^2 - \Del_n^2 + i0+}
- D_n(\nu^2, q^2) \right] +
2 C(\nu^2, q^2),
\eeq
where $D_n(\nu^2, q^2)$ is polynomial in $\nu^2$.
The terms in eq.~\dtqqtoo\ proportional to $W_n(q^2)$ depend only on form
factors, and we will refer to the contribution of these terms as the
``form-factor'' contribution.
We refer to the remaining terms as ``contact'' contributions.
(The first term in the sum over $n$ is exactly what one would write for the
\rhs\ of an {\it unsubtracted} fixed-$q^2$ dispersion relation.
As mentioned in section 2, the fixed-$q^2$ dispersion relation requires
subtraction, and so we expect the two terms in the sum on $n$ to separately
diverge, while the total sum is finite.
For this reason, we do not combine $\sum_n D_n$ with $C$.)

The \emd\ can therefore be written
\eq
\label\ourdm
\Del M_{qq} = e^2 \myint \frac{d^4 q_E}{(2\pi)^4}\, \frac 1{q^2_E}
\left\{ \sum_n \left[ \frac{\Del_n W_n(-q_E^2)}{(q_{E4})^2 + \Del_n^2} +
\avg{D_n}(q_E^2) \right] + \avg{C}(q_E^2) \right\},
\eeq
where we have continued the integral to Euclidean momentum space.
(Recall that $\avg\cdot$ denotes the 4-dimensional angular average.)

In the large-$N$ limit, the functions $C(\nu^2, q^2)$ and $D_n(\nu^2, q^2)$
have the form
\eq
C(\nu^2, q^2),\ D_n(\nu^2, q^2) \sim \sum
\frac{\hbox{polynomial in $\nu^2$, $q^2$}}{\hbox{polynomial in $q^2$}},
\eeq
where the polynomials in the denominator come from the vector-meson propagators
in fig.~1.
Therefore, after angular averaging in Euclidean space, we have
\eq
\label\ccoeff
\avg{C}(q_E^2) = \sum_{r,s} \left[
\frac{C^{(0)}_{rs} + C^{(1)}_{rs} q_E^2}
{(q_E^2 + m_{0r}^2) (q_E^2 + m_{1s}^2)} + P^{(C)}_{rs}(q_E^2) \right],
\eeq
\eq
\label\dcoeff
\avg{D_n}(q_E^2) = \sum_{r,s} \left[
\frac{D^{(0)}_{nrs} + D^{(1)}_{nrs} q_E^2}
{(q_E^2 + m_{0r}^2) (q_E^2 + m_{1s}^2)} + P^{(D)}_{nrs}(q_E^2) \right],
\eeq
where the sum over $r$, $s$ runs over the vector mesons that appear in the
propagators in fig.~1;
these have isospin 0 and 1, respectively.
(The contributions from terms that do not have two vector-meson poles can be
written in this form by choosing the polynomials in the numerator to cancel
the pole factors appearing in the denominator.)
In eqs.~\ccoeff\ and \dcoeff, $C^{(0)}, \ldots, D^{(1)}$ are constants, and
$P^{(C)}$ and $P^{(D)}$ are polynomials in $q_E^2$.

The unknown quantities on the \rhs s of eqs.~\ccoeff\ and \dcoeff\ determine
the contact contribution to the \emd.
Our strategy will be to restrict these quantities by imposing various
consistency conditions arising from the definition of $\Del T_{qq}$ in QCD.
These consistency conditions will involve the contact contribution alone
because the form factor contribution will satisfy the consistency conditions by
itself.

We first note that the integral in eq.~\ourdm\ must converge (as shown in
section 2).
It is expected on very general grounds that the form factors contributing to
$W_n(q^2)$ fall off at large $q^2$ (see below), so that for each $n$ the
form-factor contribution in eq.~\ourdm\ is convergent.
On the other hand, for general values of the coefficients in eqs.~\ccoeff\ and
\dcoeff, the contact contribution will diverge in the ultraviolet, so the
condition that the integral for $\Del M$ is finite will constrain the contact
contribution.
If we impose a cutoff $\Lam$ on the photon momentum (note that this is gauge
invariant), the divergences then have the simple forms $\Lam^{2k}$
($k = 1, 2, \ldots$) and $\log \Lam^2$.
Demanding that the coefficients of these divergences separately vanish gives
the ``ultraviolet'' sum rules
\eq
\label\trivsum
\sum_{r,s} P^{(C)}_{rs}(q_E^2) + \sum_n \sum_{r,s} P_{nrs}^{(D)}(q_E^2) = 0,
\eeq
\eq
\label\uvsum
\sum_{r,s} C^{(1)}_{rs} + \sum_n \sum_{r,s} D^{(1)}_{nrs} = 0.
\eeq
Note that eq.~\trivsum\ simply means that the polynomials $P^{(C)}$ and
$P^{(D)}$ do not contribute to $\Del M$.
To derive these sum rules, it is crucial that the meson vertices are
polynomial in momenta, so that there is a maximum possible power divergence
$\Lam^{2k}$.
If arbitrarily high powers of $\Lam^2$ were possible, then we would not be able
to unambiguously separate the logarithmic and power divergences, since an
infinite series in $\Lam^2$ may behave asymptotically like
$\ln\Lam^2$.\footnote{$^*$}
{If the vertices that contribute to $D_n$ are polynomials of higher degree as
$n$ becomes large, the integral in eq.~\ourdm\ can give rise to arbitrarily
high powers of $\Lam^2$.
However, if we make some reasonable assumptions about the asymptotic properties
of correlation functions in QCD, this can be shown not to occur.}

Next, we note that as $q \to 0$, the form factors simply measure the
appropriate charge, so from eq.~\wndata\ we have
\eq
W_n(q^2) \to (\scr Q_u^2 - \scr Q_d^2) \del_{nP}
\quad \hbox{as}\ q \to 0.
\eeq
Substituting into eq.~\dtqqtoo, we obtain
\eq
\Del T_{qq}(p,q) \to 2 \pi i (\scr Q_u^2 - \scr Q_d^2) \del(q_0)
+ 2 \sum_n D_n(0, 0) + 2 C(0, 0)
\quad \hbox{as}\ q \to 0.
\eeq
(In the first term, we have used the identity
$1 / (q_0 + i0+) + (q \to -q) = -2\pi i\del(q_0)$
to rewrite the $P$ propagator.)
The first term by itself is the correct result as $q \to 0$, as can be seen by
writing a low-energy effective theory in which only the $P$ and the $P^*$
appear.
Taking the ultraviolet sum rule eq.~\trivsum\ into account, we obtain the
``infrared'' sum rule
\eq
\label\irsum
\sum_{r,s} \frac{C^{(0)}_{rs}}{m_{0r}^2 m_{1s}^2}
+ \sum_n \sum_{r,s} \frac{D^{(0)}_{nrs}}{m_{0r}^2 m_{1s}^2} = 0.
\eeq

For the expert, we note that the sum rules in eqs.~\uvsum\ and \irsum\ are
exact analogs of the Weinberg sum rules used in the calculation of the \pid,
in the sense that the Weinberg sum rules can be derived from identical
considerations applied to the vacuum correlation function that appears in that
calculation \newpi.

So far, all of the approximations we have made have been controlled, \ie\ they
become arbitrarily accurate as some parameters of the underlying theory
approach limiting values ($N \gg 1$ and $m_Q \gg \Lam_{\rm QCD}$).
The result in eq.~\ourdm\ is an infinite sum over one-particle states with
infinitely many unknown parameters, even after the sum rules have been imposed.
However, we know from the fixed-$\qv^2$ dispersion relation discussed in
section 2 that intermediate states with large mass do not contribute
significantly to $\Del M$.
We do not know how many terms in eq.~\ourdm\ are needed to get a good
approximation to the sum, since there is no known small parameter controlling
the convergence of the series.

In order to make progress, we simply {\it assume} that the sum is well
approximated by the first non-trivial term, \ie\ that it is a good
approximation to retaining only the mimimal set of intermediate states that
gives a consistent description of the matrix elements that appear in the
calculation.
In particular, it is crucial that the matrix elements have the correct
ultraviolet and infrared behavior to satisfy the sum rules derived above.

Clearly, we must include both the $P$ and the $P^*$ as intermediate states,
since they become degenerate in the heavy-quark limit.
To see what other states we must include, we consider the matrix elements
\eqa
\label\fffdef
\bra{P(p)} J^\mu_{q0,1}(0) \ket{P(p + q)} &= F_{0,1}(-q^2)\, v^\mu, \eol
\label\ffgdef
\bra{P(p)} J^\mu_{q0,1}(0) \ket{P^*(p + q, \ep)} &= i G_{0,1}(-q^2)\,
\ep^{\mu\nu\rho\sig} v_\nu q_\rho \ep_\sig, \eeol
\eeq
which determine the form factor contribution if we include no heavy mesons
other than the $P$ and $P^*$.
In the appendix, we use the constituent counting rules
\ref\constit{S. J. Brodsky and G. Farrar, \PRL{31}{1153}{1973};
\PRD{11}{1309}{1975};
S. J. Brodsky and G. P. Lepage, \PRD{22}{2157}{1980}.
For a recent review, see S. J. Brodsky and G. P. Lepage in {\it Perturbative
Quantum Chromodynamics}, edited by A. H. Mueller (World Scientific, 1989).}\
to show that the form factors $F$ and $G$ fall off as $\sim 1/q^4$ for large
spacelike momentum transfer $q$.
In the large-$N$ limit, the falloff of these form factors is due to the
presence of single vector-meson intermediate states (see fig.~2).
Since a single vector-meson pole can fall off at most as $\sim 1/q^2$, we see
that the asymptotic behavior requires cancelation between at least two
different vector-meson intermediate states.
The minimal set of vector-meson states we must consider therefore consists of
the lightest isospin-1 vector mesons $\rho$ and $\rho'$ and the corresponding
isospin-0 mesons $\om$ and $\om'$.
{}From the behavior of the form factors at small and large $q$, we can write
\eqa
\label\fff
F_{0,1}(q_E^2) &\simeq -\frac{\scr Q_u \pm \scr Q_d}2\,
\frac 1{(1 + q_E^2 / m^2_{\rho})(1 + q_E^2 / m^2_{\rho'})}, \eol
\label\ffg
G_{0,1}(q_E^2) &\simeq -\frac{\scr Q_u \pm \scr Q_d}2\,
\frac{\beta/2}{(1 + q_E^2 / m^2_{\rho})(1 + q_E^2 / m^2_{\rho'})}, \eeol
\eeq
where we have used the fact that $m_\rho = m_\om$ and $m_{\rho'} = m_{\om'}$ in
the large-$N$ limit.
(Numerically, $m_\om - m_\rho = 14 \MeV$.)
Here, $\beta$ is a strong-interaction matrix element that can be determined
from the $P^* \to P \gam$ decay rate.
It is important to note that eqs.~\fff\ and \ffg\ are to be understood as
valid only for the small $q_E^2$ that dominate in the integral that determines
$\Del M$.

We therefore truncate the intermediate states by keeping only the contributions
from the $P$, $P^*$, $\rho$, $\om$, $\rho'$, and $\om'$.
The contribution to $\Del M_{qq}$ from the form factor contribution to
eq.~\ourdm\ can be directly evaluated using eqs.~\wndata\ and \fffdef--\ffg.
We have used the language of dispersion theory to make the connection to
physical matrix elements explicit, but our final result can be stated very
simply in terms of Feynman graphs:
we evaluate the one-loop electromagnetic self-energy for the heavy meson with
momentum-dependent photon couplings as given in eqs.~\fff\ and \ffg.
The result is
\eq
\Del M^{\rm form\,factor}_{qq} \simeq (\scr Q_u^2 - \scr Q_d^2) \al
\left[ \frac{m_\rho}4\, \frac{1 + 3x + x^2}{(1 + x)^3}
- \frac{\beta^2 m_\rho^3}{8}\, \frac{1}{(1 + x)^3} \right],
\eeq
where
\eq
x \equiv \frac{m_\rho}{m_{\rho'}} = 0.53.
\eeq
(We use the symbol ``$\simeq$'' to denote statements that are valid only with
the truncation of states discussed above.)
Numerically,
\eq
\label\qqnum
\Del M^{\rm form\,factor}_{qq} \simeq
+0.38 - 0.039 \left( \frac{\beta}{1 \GeV^{-1}} \right)^2 \MeV.
\eeq
The value of $\beta$ has not yet been measured directly.
(There is a weak upper bound that results from using $SU(3)$ symmetry applied
to the decays $D^* \to D \gam$ and $D^* \to D \pi$
\ref\betabound{J. F. Amundson \etal, \PLB{296}{415}{1992};
P. Cho and H. Georgi, \PLB{296}{408}{1992}.}.
Hadronic models tend to give values of $D^* \to D \gam$ and $D^* \to D \pi$
consistent with $\beta$ near $1 \GeV^{-1}$
\ref\betamodel{See \eg\ B. Holdom, S. Jaimungal, R. Lewis, and M. Sutherland,
UTPT-94-25, hep-ph/9410324;
V. M. Belyaev, V. M. Braun, A. Khodjamirian, and R. R\"uckl, MPI-PHT/94-62,
hep-ph/9410280.}.)
We will see that the contribution in eq.~\qqnum\ is numerically small
compared to $\Del M_{Qq}$ computed in the next subsection, and the uncertainty
in the value of $\beta$ will not be very important for our final results.
This issue will be analyzed in greater detail in a subsequent paper \us.

We now turn to the contact contribution in eq.~\dtqqtoo.
The ultraviolet sum rule eq.~\uvsum\ allows us to parameterize the contact
contribution as
\eq
\label\contactparam
\avg{\Del T^{\rm contact}}(q_E^2) \simeq
(\scr Q_u^2 - \scr Q_d^2) m_\rho^3 \left[
\frac{A_{\rho\rho}}{(q_E^2 + m_\rho^2)^2}
+ \frac{A_{\rho\rho'}}{(q_E^2 + m_\rho^2)(q_E^2 + m_{\rho'}^2)}
+ \frac{A_{\rho'\rho'}}{(q_E^2 + m_{\rho'}^2)^2} \right],
\eeq
where the $A$'s are linear combinations of the coefficients $C$ and $D$
defined in eq.~\ourdm.
We have normalized the $A$'s so that they are $\sim 1$ if we identify $m_\rho$
with the hadronic scale that appears in these matrix elements.
Imposing the infrared sum rule eq.~\irsum, we obtain
\eq
A_{\rho\rho} \simeq -x^2 A_{\rho\rho'} - x^4 A_{\rho'\rho'}.
\eeq
It is now straightforward to compute the contact contribution to the \emd.
We obtain
\eqa
\Del M_{qq}^{\rm contact} &\simeq -(\scr Q_u^2 - \scr Q_d^2)
\frac{\al m_\rho}{8\pi} x^2
\left[ A_{\rho\rho'} \frac{1 - x^2 + \ln x^2}{1 - x^2}
+ A_{\rho'\rho'} (1 - x^2) \right] \eol
&= -0.02\, A_{\rho\rho'} + 0.02\, A_{\rho'\rho'} \MeV. \eeol
\eeq
We see that for any reasonable value of the $A$'s, this contribution is
numerically negligible, and $\Del M_{qq}$ is given by eq.~\qqnum\ to a good
approximation.

Heavy-quark symmetry equates the electromagnetic mass difference for the
$P$ to that of the $P^*$.
We have (in an obvious notation)
\eq
\Del M^*_{qq} = \Del M_{qq}.
\eeq

Before continuing the discussion, we note that our final result is rather
similar to the prescription given long ago for computing hadron \emd s
\oldbarry, which was essentially to evaluate the electromagnetic self-energy
with the photon vertices replaced by momentum-dependent form factors.
However, we emphasize that this prescription was never justified in any
satisfactory way.
In particular, the form factors appearing used in this prescription must be
evaluated off shell in order to compute the \emd.
Clearly, there are infinitely many ways to continue a form factor off shell,
and these will give different final results.
(In our formalism, this ambiguity is contained in the term $\avg{D_n}$ defined
in eq.~\ourdm.)
Also, neglecting the contact contributions (contained in $C$ in our
formalism) was never justified.

The use of the large-$N$ limit and the truncation of intermediate states may
seem like rather drastic approximations.
However, we note that precisely analogous approximations are used in the
classic result \oldpi\ for the \pid:
the \pid\ is written in terms of vacuum correlation functions, which are then
saturated with the appropriate lowest-lying 1-particle states in the large-$N$
limit (in this case, the $\rho$ and the $a_1$ vector mesons).
In this way, one can obtain a formula similar to the ones obtained here for
the heavy meson \emd s \oldpi\newpi:
\eq
m_{\pi^+} - m_{\pi^0} \simeq \frac{3\al}{8\pi m_\pi}\,
\frac{m_\rho^2}{1 - m_\rho^2 / m_a^2}\,
\ln\frac{m_a^2}{m_\rho^2}
\simeq 5.9 \MeV.
\eeq
This is within $30\%$ of the experimental value $4.6 \MeV$.
We therefore adopt this as an estimate of the size of the error from the
{\it combined} approximation of the large-$N$ limit and the truncation of
intermediate states.

As stated in the introduction, the discussion that follows consists mainly
of repeated applications of the principles described above to compute the
remaining contributions to $\Del M$.
The weary reader interested mainly in the bottom line is invited to skip to
section 5, which summarizes the main results and gives our conclusions.

\subsection{$\Del M_{Qq}$}
The computation of $\Del M_{Qq}$ can be carried out following the same
arguments given in subsection 3.1, so we will be brief.
We can write
\eq
\Del T_{Qq}(\nu^2, q^2) =
-2 \sum_n \left[ \frac{\Del_n W_n(q^2)}{\nu^2 - \Del_n^2 + i0+}
- D_n(\nu^2, q^2) \right] + 2 C(\nu^2, q^2),
\eeq
where the functions $W_n$, $D_n$, and $C$ are defined as in eq.~\dtqqtoo,
but with $J_Q^\mu$ replacing $J_{q0}^\mu$.

The main difference between $\Del T_{Qq}$ and $\Del T_{qq}$ arises because
the form factors of $J_Q^\mu$ are constants in the heavy-quark limit
\aref{N. Isgur and M. B. Wise, \PLB{232}{113}{1989}; \PLB{237}{527}{1990}.}:
\eqa
\label\fqf
\bra{P(p)} J^\mu_Q(0) \ket{P(p + q)} &= \scr Q_Q v^\mu, \eol
\bra{P(p)} J^\mu_Q(0) \ket{P^*(p + q, \ep)} &= 0. \eeol
\eeq
The fact that the form factors are constants even for large momenta can also be
understood from the point of view of the constituent counting rules \constit,
since graphs in which the ``hard'' momentum flows directly into the
heavy-quark line do not fall off at large $|q^2|$ (see the appendix).
{}From the point of view of the large-$N$ meson theory, we can understand this
in the following way:
the intermediate meson states that contribute to the form factor for
$J_Q^\mu$ in the large-$N$ limit are $Q\mybar Q$ states, and so have mass
of order $2 m_Q$.
These states are integrated out of the effective theory
, and their effects
are correctly taken into account by local interactions.
The form factors are therefore pure polynomials in $q^2$ and $\nu$ in the
heavy-quark limit.
Because we do not expect the form factors to grow, they must be constants, and
the constants are fixed by the value of the form factors at $q = 0$.
(Similar arguments are made in
ref.~\aref{B.~Grinstein and P.~F.~Mende, \NPB{425}{451}{1994}.}.)

{}From eq.~\fqf, we then have
\eqa
\label\avgcdef
\avg{C}(q_E^2) &= \sum_s \left[
\frac{C^{(0)}_s}{q_E^2 + m_{1s}^2} + P_s^{(C)}(q_E^2) \right], \eol
\label\avgddef
\avg{D_n}(q_E^2) &= \sum_s \left[
\frac{D^{(0)}_{ns}}{q_E^2 + m_{1s}^2} + P_{ns}^{(D)}(q_E^2) \right], \eeol
\eeq
where the sum over $s$ is over isospin-1 vector mesons that appear in the
propagators in figs.~1 and 2.
As before, $C^{(0)}_s$ and $D^{(0)}_{ns}$ are constants, while $P^{(C)}$ and
$P^{(D)}$ are polynomials in $q_E^2$.

We now write ultraviolet and infrared sum rules using the same reasoning used
in the previous subsection.
We have
\eqa
\sum_s P^{(C)}_s(q_E^2) + \sum_n \sum_s P^{(D)}_{ns}(q_E^2) &= 0, \eol
\sum_s C^{(0)}_s  + \sum_n \sum_s D^{(0)}_{ns} &= 0, \eol
\label\irsum
\sum_s \frac{C^{(0)}_s}{m_{1s}^2} + \sum_n \sum_s \frac{D^{(0)}_{ns}}
{m_{1s}^2} &= 0. \eeol
\eeq
As before, we approximate $\Del M_{Qq}$ by keeping only the contributions of
the lowest-lying states;
in this case, only the $P$ heavy meson state and the $\rho$ and $\rho'$ vector
mesons are required.

It is not hard to see that the contact contribution is forced to vanish
identically when the sum rules are saturated by $\rho$ and $\rho'$ vector
mesons.
The form factor for the light-quark current is given in eq.~\fff\ in the
approximation we are making.
Substituting this and the heavy-quark current form factor in eq.~\fqf\ into
an expression for $\Del M_{Qq}$ analogous to eq.~\ourdm, we obtain
\eqa
\Del M_{Qq} &\simeq
- \scr Q_Q (\scr Q_u - \scr Q_d) \frac{\al m_\rho}{1 + x} \eol
\label\Qqnum
&= \cases{ -2.5 \MeV & for $Q = c$, \cr
+1.2 \MeV & for $Q = b$. \cr} \eeol
\eeq
We see that this contribution numerically dominates the light-quark current
contribution of eq.~\qqnum.
Once again, heavy-quark symmetry gives
\eq
\label\Qqsrel
\Del M^*_{Qq} = \Del M_{Qq}.
\eeq

\section{$1/m_Q$ Corrections}
In this section, we consider the $1/m_Q$ corrections to the results obtained
above, still working in the large-$N$ limit.
For mesons containing a $c$ quark, these corrections are expected to be
substantial, and so it is important to estimate them.

\subsection{$\Del M_{qq}$}
To include $O(1/m_Q)$ effects, we must make two types of changes to the
formulas of the previous section.
First, the meson vertices and the masses change by $O(1/m_Q)$ effects.
Second, the form of the heavy-meson propagators is modified.
This modification can be thought of as a recoil correction: it is completely
kinematical in origin, and hence determined with no dynamical input.
Formally, this can be thought of as writing a non-relativistic expansion of
the fully relativistic propagator or (more correctly) an expression of
reparameterization invariance \reparam.
The result is that eq.~\ourdm\ in the previous section is replaced with
\eq
\label\ourdmm
\Del M_{qq} = e^2 \myint \frac{d^4 q_E}{(2\pi)^4}\, \frac 1{q^2_E}
\left\{ \sum_n \left[ \frac{(\Del_n + \del\Del_n(q_E))W_n(-q_E^2)}
{(q_{E4})^2 + \Del_n^2} +
\avg{D_n}(q_E^2) \right] + \avg{C}(q_E^2) \right\},
\eeq
where
\eq
\del\Del_n(q_E) = \frac{q_E^2 + \Del_n^2}{2M}
\left(1 - \frac{2\Del_n^2}{q_{E4}^2 + \Del_n^2} \right).
\eeq
The functions $\avg{C}$ and $\avg{D_n}$ are given by formulas identical to
eqs.~\ccoeff\ and \dcoeff, except that the coefficients are now given in
a power series in $1/m_Q$:
\eq
\label\mqexpand
\eqalign{
C^{(0)}_{rs} &=
C^{(0,0)}_{rs} + C^{(0,1)}_{rs}  + O(1/m_Q^2), \cr
\Del M_{qq} &=
\Del M_{qq}^{(0)} + \Del M_{qq}^{(1)} + O(1/m_Q^2), \cr}
\eeq
\etc., where $C^{(0,k)}$, $\Del M_{qq}^{(k)},\ldots$ parameterize the
$O(1/m_Q^k)$ corrections.

We obtain ultraviolet sum rules by imposing the condition that the integral
for $\Del M_{qq}$ converges order by order in $1/m_Q$.
Because the form factors fall off as $\sim 1/q^3$ (see the appendix), the
form-factor contribution for each $n$ converges by itself, and the sum rules
again constrain the contact contribution alone:
\eq
\label\mqtrivsum
\sum_{r,s} P^{(C)}_{rs}(q_E^2) + \sum_n \sum_{r,s} P_{nrs}^{(D)}(q_E^2) = 0,
\eeq
\eq
\label\mquvsum
\sum_{r,s} C^{(1,j)}_{rs} + \sum_n \sum_{r,s} D^{(1,j)}_{nrs} = 0,
\quad j = 0,1.
\eeq

To get the infrared sum rules, we note that the recoil corrections do not
affect the leading behavior of the integrand as $q_E \to 0$.
However, there is a new feature that enters at $O(1/m_Q)$ because
\eq
\Del T(p, q) \to  (\scr Q_u^2 - \scr Q_d^2)
\left[2 \pi i \del(q_0) + \frac 4M \right]
\quad\hbox{as}\ q \to 0.
\eeq
This results by writing a low-energy effective theory where only the $P$ and
the $P^*$ appear.
In this theory, the ``extra'' $1/M$ term arises from the term
\eq
\label\extra
\del\scr L_{\rm eff} = -\frac{1}{2M} P^\dagger D^2 P,
\eeq
where the coefficient is fixed exactly by reparameterization invariance.
(Other interactions that arise at order $1 / m_Q$ vanish as $q \to 0$.)
Because of this, the infrared sum rules are
\eq
\label\newir
\sum_{r,s} \frac{C^{(0,0)}_{rs}}{m_{0r}^2 m_{1s}^2}
+ \sum_n \sum_{r,s} \frac{D^{(0,0)}_{nrs}}{m_{0r}^2 m_{1s}^2} = 0.
\eeq
\eq
\label\newirtoo
\sum_{r,s} \frac{C^{(0,1)}_{rs}}{m_{0r}^2 m_{1s}^2}
+ \sum_n \sum_{r,s} \frac{D^{(0,1)}_{nrs}}{m_{0r}^2 m_{1s}^2}
= \frac{2}{M} (\scr Q_u^2 - \scr Q_d^2)
\eeq
We have once again used the fact that the form factor contribution coming from
$W_n(q^2)$ already has the correct infrared behavior, except for
the term in eq.~\extra.

These sum rules can be saturated by the same set of states as in the
previous section: $\rho$, $\om$, $\rho'$, $\om'$, $P$, and $P^*$.
The form factor contribution is convergent by itself, and is given by
\eq
\eqalign{
\Del M_{qq}^{(1)\,{\rm form\,factor}} \simeq &
-\frac{(\scr Q_u^2 - \scr Q_d^2) \al m_\rho^2}{4\pi M}\,
\frac{1 + 2 x^2\ln x^2 - x^4}{(1 - x^2)^3} \cr
&\quad + \frac{3 (\scr Q_u^2 - \scr Q_d^2) \al\beta^2 m_\rho^2}{8\pi}
\biggl[ (M^* - M)\, \frac{1 + 2 x^2\ln x^2 - x^4}{(1 - x^2)^3} \cr
&\qquad\qquad\qquad\qquad\qquad\qquad\quad
- \frac{m_\rho^2}{2M}\,\frac{2(1 - x^2) + (1 + x^2)\ln x^2}
{(1 - x^2)^3} \biggr]. \cr}
\eeq
Using the infrared sum rule, the contact contribution can be written
(compare to eq.~\contactparam)
\eq
\eqalign{
\avg{\Del T^{(1)\,\rm contact}}(q_E^2) &\simeq
\frac{4}{M}\,\frac{ (\scr Q_u^2 - \scr Q_d^2) m_\rho^4}
{(q_E^2 + m_\rho^2)^2} \cr
&\quad + (\scr Q_u^2 - \scr Q_d^2) m_\rho^3 \left[
\frac{A^{(1)}_{\rho\rho}}{(q_E^2 + m_\rho^2)^2}
+ \frac{A^{(1)}_{\rho\rho'}}{(q_E^2 + m_\rho^2)(q_E^2 + m_{\rho'}^2)}
+ \frac{A^{(1)}_{\rho'\rho'}}{(q_E^2 + m_{\rho'}^2)^2} \right]. \cr}
\eeq
The constants $A^{(1)} \sim m_\rho / M$ parameterize the $O(1/m_Q)$
corrections to the unknown contact contributions defined in eq.~\contactparam.
The ultraviolet sum rule then gives
\eq
A^{(1)}_{\rho\rho} \simeq -x^2 A^{(1)}_{\rho\rho'}
- x^4 A^{(1)}_{\rho'\rho'},
\eeq
and we obtain
\eq
\eqalign{
\Del M_{qq}^{(1)\,{\rm contact}} & \simeq
\frac{(\scr Q_u^2 - \scr Q_d^2) \al m_\rho}{8\pi M} \cr
&\qquad
- \frac{(\scr Q_u^2 - \scr Q_d^2) \al m_\rho}{8\pi} x^2 \left[
A^{(1)}_{\rho\rho'} \frac{1 - x^2 + \ln x^2}{1 - x^2}
+ A^{(1)}_{\rho'\rho'} (1 - x^2) \right]. \cr}
\eeq
We find that the corrections that depend on the $A$'s are numerically
negligible for reasonable values of $A^{(1)}$.
Our result is therefore
\eq
\label\qqnumcorr
\Del M_{qq}^{(1)} \simeq \cases{\displaystyle
0.088 + 0.028 \left(\frac{\beta}{1 \GeV^{-1}} \right)^2_{\vphantom{{}_\j}} \MeV
& for $Q = c$, \cr\displaystyle
0.031 + 0.0094 \left(\frac{\beta}{1 \GeV^{-1}} \right)^2 \MeV
& for $Q = b$. \cr}
\eeq
Comparing to the lowest-order results in eq.~\qqnum, we see that the $1/m_Q$
corrections are $35\%$ for the $c$ system and $10\%$ for the $b$ system for
$\beta = 1 \GeV^{-1}$.

At order $1/m_Q$ the  $P^*$ \emd\ is no longer equal to that of the $P$.
Applying the same method, we obtain the isospin-violating
hyperfine splitting due to the light quark charges,
\eq
\eqalign{
\Del M^*_{qq} - \Del M_{qq}
&\simeq -\frac{(\scr Q_u^2 - \scr Q_d^2) \al \beta^2 m_\rho^2}{2\pi}\,
(M^* - M)\,
\frac{1 + 2 x^2 \ln x^2 - x^4}{(1 - x^2)^3} \cr
&\qquad - \frac{(\scr Q_u^2 - \scr Q_d^2) \al \beta m_\rho^3}{6}\,
(\beta' - \beta)\,
\frac{1}{(1 + x)^3} \cr}
\eeq
where $\beta'$ measures the strength of a $P^*$--$P^*$--$\gam$ coupling.
Heavy-quark symmetry gives $\beta' = \beta + O(1/m_Q)$, but we do not have
any experimental information about the difference $\beta' - \beta$.
We therefore obtain
\eq
\label\qqstarnumcorr
\Del M^*_{qq} - \Del M_{qq} \simeq \cases{\displaystyle
-0.16 - 0.015
\left(\frac{\beta}{1 \GeV^{-1}} \right)^2_{\vphantom{{}_\j}}
\left( \frac{\beta' - \beta}{0.3\,\, \beta} \right) \MeV &
for $Q = c$, \cr\displaystyle
-0.054 - 0.0052
\left(\frac{\beta}{1 \GeV^{-1}} \right)^{2\vphantom{{}^A}}
\left( \frac{\beta' - \beta}{0.1\,\, \beta} \right) \MeV &
for $Q = b$, \cr}
\eeq
where we have normalized the value of $\beta' - \beta$ to a representative
value.
Comparing to eq.~\qqnum, we see that the $1/m_Q$ correction to $\Del M_{qq}^*$
is $20\%$ for the $c$ system and $5\%$ for the $b$ system for the values
of $\beta$ and $\beta'$ used to normalize these expressions.

\subsection{$\Del M_{Qq}$}
When we include the $O(1/m_Q)$ corrections, $\Del M_{Qq}$ is given by a
formula just like eq.~\ourdmm, but with $J_Q^\mu$ replacing $J_{q0}^\mu$.
The functions $\avg C$ and $\avg{D_n}$ are given by formulas like
eqs.~\avgcdef\ and \avgddef, but with the coefficients expanded in powers
of $1/m_Q$, as in the last section (see eq.~\mqexpand).
Using reasoning similar to that of the previous subsection, we obtain
ultraviolet sum rules of the same form as eqs.~\mqtrivsum\ and \mquvsum,
and infrared sum rules of the same form as eqs.~\newir\ and \newirtoo.
The contact contribution is uniquely determined when the sum rules are
imposed.
Performing the necessary computations, we obtain
\eq
\Del M_{Qq}^{(1)} \simeq
\frac{\scr Q_Q (\scr Q_u - \scr Q_d) \al m_\rho^2}{2\pi M}\,
\frac{\ln x^2}{1 - x^2}
+ \frac{\scr Q_Q (\scr Q_u - \scr Q_d) \al \beta\beta_Q m_\rho^3}{2}\,
\frac{1}{x(1 + x)},
\eeq
where $\beta_Q$ is the coupling of the heavy-quark current to $P$ and $P^*$,
normalized like $\beta$ in eq.~\ffg.
In the heavy-quark limit,
\eq
\label\betaQ
\beta_Q = \frac{1}{M}\, \left[1 + O(\al_{\rm s}(M)) \right].
\eeq
The corrections are expected to be of order $\al_{\rm s}(m_Q) / \pi$, which
is $0.15$ for the $c$ quark and $0.08$ for the $b$ quark.
Neglecting these corrections, we obtain
\eq
\label\Qqnumcorr
\Del M_{Qq}^{(1)} = \cases{\displaystyle
-0.44 + 0.73 \left( \frac\beta{1 \GeV^{-1}} \right)_{\vphantom{{}_\j}}
\MeV & for $Q = c$, \cr\displaystyle
+0.077 - 0.13 \left( \frac\beta{1 \GeV^{-1}} \right) \MeV & for $Q = b$. \cr}
\eeq
Comparing to eq.~\Qqnum, we see that this is a $10\%$ correction for the $c$
system and a $15\%$ correction for the $b$ system for $\beta = 1 \GeV^{-1}$.
(The small size of the corrections for the $c$ system is a result of
cancelations that depend on the value of $\beta$.)

Similarly, we can compute the $P^*$ isospin splittings:
\eq
\Del M_{Qq}^* - \Del M_{Qq} \simeq
-\frac{2 \scr Q_Q(\scr Q_u - \scr Q_d) \al \beta \beta_Q m_{\rho}^3}{3}\,
\frac{1}{x(1 + x)},
\eeq
which gives (again using eq.~\betaQ)
\eq
\label\Qqstarnumcorr
\Del M_{Qq}^* - \Del M_{Qq} \simeq \cases{\displaystyle
-0.99 \left( \frac\beta{1 \GeV^{-1}} \right)_{\vphantom{{}_\j}}
\MeV & for $Q = c$, \cr\displaystyle
+0.17 \left( \frac\beta{1 \GeV^{-1}} \right) \MeV & for $Q = b$. \cr}
\eeq

\section{Summary and Conclusions}
We now summarize our results.
We have computed the heavy-meson \emd s by working in the large-$N$ limit,
where the \emd s are given by a convergent sum over an infinite number of
1-particle intermediate states.
We obtained sum rules relating the matrix elements appearing in this sum that
enforce the correct ultraviolet and infrared behavior on the electromagnetic
amplitudes that appear in the calculation.
All of this is a rigorous consequence of QCD (in the large-$N$ limit);
however, in order to get numerical results, we truncated the infinite sum by
keeping the smallest number of intermediate states that are capable of giving
a consistent description of the matrix elements which appear in the sum.
We argued that these approximations are similar to the ones made in the
classic calculation of the \pid, which works to $30\%$.
Making these approximations, we find that the numerically dominant contribution
comes from the heavy-quark current, and we obtain
(see eqs.~\qqnum, \Qqnum, \qqnumcorr, and \Qqnumcorr)
\eqa
(M_{D^0} - M_{D^+})^{\rm EM} &\simeq
\biggl[ -2.4 + 0.74 \left( \frac{\beta}{1 \GeV^{-1}} \right)
-0.012 \left( \frac{\beta}{1 \GeV^{-1}} \right)^2 \biggr] \MeV \eolnn
&\qquad + O(1/m_c^2), \eol
(M_{B^+} - M_{B^0})^{\rm EM} &\simeq
\biggl[ +1.7 - 0.13 \left( \frac{\beta}{1 \GeV^{-1}} \right)
-0.03 \left( \frac{\beta}{1 \GeV^{-1}} \right)^2 \biggr] \MeV \eolnn
&\qquad + O(1/m_b^2). \eeol
\eeq
Here, $\beta$ is a matrix element that measures the strength of the
$P^*$--$P$--$\gam$ coupling (see eqs.~\ffgdef\ and \ffg).
The coefficients of the terms linear in $\beta$ are $O(1/m_Q)$ and have
perturbative QCD corrections of order $\al_{\rm s}(m_Q)$ (see eq.~\betaQ).
Similar results are also obtained for the vector mesons.
Heavy quark symmetry gives $\Del M^* = \Del M + O(1/m_Q)$, and we obtain
(see eqs.~\qqstarnumcorr\ and \Qqstarnumcorr)
\eqa
(M_{D^{*0}} - M_{D^{*+}})^{\rm EM} &\simeq
(M_{D^0} - M_{D^+})^{\rm EM} \eolnn
&\quad + \biggl[ - 0.16 - 0.99 \left( \frac{\beta}{1 \GeV^{-1}} \right)
- 0.015 \left(\frac{\beta}{1 \GeV^{-1}} \right)^2
\left( \frac{\beta' - \beta}{0.3\,\, \beta} \right) \biggr] \MeV \eolnn
&\quad + O(1 / m_c^2) \eol
(M_{B^{*+}} - M_{B^{*0}})^{\rm EM} &\simeq
(M_{B^+} - M_{B^0})^{\rm EM} \eolnn
&\quad + \biggl[ -0.054 + 0.17 \left( \frac{\beta}{1 \GeV^{-1}} \right)
- 0.0052 \left(\frac{\beta}{1 \GeV^{-1}} \right)^2
\left( \frac{\beta' - \beta}{0.1\,\, \beta} \right) \biggr] \MeV \eolnn
&\quad + O(1 / m_b^2), \eeol
\eeq
where $\beta' = \beta + O(1/m_Q)$ is another unmeasured matrix element
(see the discussion above eq.~\qqstarnumcorr).
As above, the coefficients of the terms linear in $\beta$ are $O(1/m_Q)$ and
have corrections of order $\al_{\rm s}(m_Q)$.

For $\beta$ near $1 \GeV^{-1}$, the heavy-quark expansion appears to be
working very well for the $b$ system, and moderately well for the $c$
system.
We believe that this is sufficiently encouraging to consider the phenomenology
of these results in detail in a subsequent paper \us.
For the present, we hope that the systematic approach taken in this paper
is the starting point for further progress on this classic problem in
hadronic physics.

\section{Acknowledgements}
We would like to thank S.~J.~Brodsky, J.~F.~Donoghue, A.~Falk, R.~L.~Jaffe,
X.~Ji, R.~F.~Lebed, M.~Savage and M.~Suzuki for helpful discussions.
We also thank the Institute for Theoretical Physics at Santa Barbara for
hospitality while this work was in progress.
MAL is supported in part by DOE contract DE-AC02-76ER03069 and by NSF
grant PHY89-04035.
RS is supported in part by NSF Grant NSF PHY-92-18167.
\vfill\eject

\appendix{Ultraviolet Behavior of Heavy-Meson Form Factors}
In this appendix, we consider the ultraviolet behavior of matrix elements of
the form
\eq
\label\matel
\bra{P(p)} J_q^\mu(0) \ket{n(p + q)},
\eeq
where $n$ is a heavy-light meson.
We work in the the $1/m_Q$ expansion.
This means that we consider spacelike momenta $q$ in the limit
\eq
\LQCD^2 \ll |q^2| \ll m_Q^2.
\eeq
The ultraviolet behavior of these form factors can be determined from the
constituent counting rules \constit.
The idea is that for large spacelike momentum transfer, the leading behavior
of the form factor can be obtained by power counting the hard momentum flow
through constituent Feynman diagrams with appropriate kinematics for the
initial and final state constituents.
For the case under consideration, the leading graphs come from a single hard
gluon exchange, as shown in the first two graphs in fig.~3.
Evaluated in the heavy-quark effective theory, these graphs give a contribution
of the form
\eq
{\rm fig.~3} \sim \Avg{
\frac{\mybar\psi'(k') \sla v (\sla k + \sla q) \gam^\mu \psi(k)}
{(k + q)^2 (x q)^2} +
\frac{\mybar\psi'(k') \gam^\mu (\sla k - x \sla q) \sla v \psi(k)}
{(k - x q)^2 (x q)^2} } + \cdots,
\eeq
where $x$ is the fraction of the hard momentum $q$ flowing into the heavy
quark.
The average is taken over $k$ and $x$.
The dominant contribution will come from regions of integration where almost
all of the hard momentum flows into the heavy quark (\ie, $k \sim \LQCD$ and
$x = 1 + O(\LQCD / m_Q)$).
The reason is simply that if hard momentum flows into the light quark, the
final state of the hard scattering subprocess consists of the heavy quark at
rest and a light quark with momentum much larger than $\Lambda_{\rm QCD}$;
such a state is expected to have very small overlap with a heavy meson at rest.
Therefore,
\eq
\label\oneglue
{\rm fig.~3} \sim \frac{q_\nu}{q^4}
\bigl\langle \mybar\psi'(k')
\left[ \sla v \gam^\nu \gam^\mu - \gam^\mu \gam^\nu \sla v \right]
\psi(k) \bigr\rangle
+ \frac{1}{q^4} \bigl\langle \mybar\psi'(k')
\left[ \sla v \sla k \gam^\mu + \gam^\mu \sla k \sla v \right]
\psi(k) \bigr\rangle
+ \cdots,
\eeq
and we have
\eqa
\bra{P(p)} J_q^\mu(0) \ket{P(p + q)} &\sim \frac{\LQCD^2}{q^4}, \eol
\bra{P(p)} J_q^\mu(0) \ket{P^*(p + q)} &\sim \frac{\LQCD}{q^3}, \eeol
\eeq
\etc.
For the $P$ elastic form factor, we have used the fact that rotational
symmetry implies that the first term in eq.~\oneglue\ is proportional to
$q_\nu v^\mu v^\nu = O(1/m_Q)$, since the fact that the initial and final
states have the same mass forces $q \cdot v = - q^2/2M$.
In general, it is clear that matrix elements of the form of eq.~\matel\ fall
off at least as fast as $\sim 1 / q^3$.
This is the result quoted in the main text.

A complete analysis of the form factors would include a discussion of
higher-order graphs such as the last one in fig.~3.
We will not give a detailed analysis of this issue here.
However, we expect graphs such as this to factorize, so that regions of
integration where the gluon momenta are soft can be absorbed into
``wavefunction'' corrections, while the remaining hard contributions have
the same asymptotic behavior as the contributions analyzed above (up to
logarithms).
For an example of this type of analysis, see
ref.~\ref\piform{A.~Duncan and A.~H.~Mueller, \PRD{21}{1636}{1980}}.


\listrefs
\vfill
\eject
\centerline{\bf Figure Captions}
\vskip .4in
\noindent
Fig.~1.
Contributions to the Compton amplitude $T$ in the large-$N$ limit.
The sum over $n$ runs over excited heavy mesons, while the sum over
$r$ and $s$ runs over vector mesons.
The shaded blobs on the graphs involving a sum over $n$ are heavy meson
form factors;
see fig.~2.

\vskip .4in
\noindent
Fig.~2.
Contributions to heavy-meson form factors in the large-$N$ limit.
The sum over $r$ runs over vector meson states.

\vskip .4in
\noindent
Fig.~3.
Contributions to the quark scattering amplitude used to determine the
asymptotic behavior of the light-quark current form factors.
The thick line is the heavy quark, and the curly line represents the gluon
propagator.

\bye